# Spin orbit effects in CoFeB/MgO hetereostructures with heavy metal underlayers

Jacob Torrejon[1], Junyeon Kim[2], Jaivardhan Sinha[3] and Masamitsu Hayashi[4,5*]

[1]*Unité Mixte de Physique CNRS/Thales, 1 Avenue Augustin Fresnel, 91767 Palaiseau, France*

[2]*Center for Emergent Matter Science, RIKEN, 2-1 Hirosawa, Wako, Saitama 351-0198, Japan*

[3]*Thematic Unit of Excellence on Nanodevice Technology, Department of Condensed Matter Physics and Material Sciences, S. N. Bose National Centre for Basic Sciences, Kolkata 700 106, India*

[4]*Department of Physics, The University of Tokyo, Bunkyo, Tokyo 113-0033, Japan*

[5]*National Institute for Materials Science, Tsukuba 305-0047, Japan*

We study effects originating from the strong spin orbit coupling in CoFeB/MgO heterostructures with heavy metal (HM) underlayers. The perpendicular magnetic anisotropy at the CoFeB/MgO interface, the spin Hall angle of the heavy metal layer, current induced torques and the Dzyaloshinskii-Moriya interaction at the HM/CoFeB interfaces are studied for films in which the early 5$d$ transition metals are used as the HM underlayer. We show how the choice of the HM layer influences these intricate spin orbit effects that emerge within the bulk and at interfaces of the heterostructures.

*Email: hayashi@phys.s.u-tokyo.ac.jp

1. Introduction

Spin orbit coupling plays an essential role in modern spintronics[1, 2]. The spin Hall effect[3] and the Rashba-Edelstein effect[4, 5], which enable spin current generation and/or spin accumulation, originate from the strong spin orbit coupling within the bulk of a heavy metal layer or at interfaces of different materials. The accumulated spins can diffuse into neighboring magnetic layer(s) to exert torque on the magnetic moments[6-8]. Such current induced torque allows a new approach to manipulate magnetic moments and may open pathways for device applications that were not possible just with the conventional spin transfer torque. Recent reports have shown that indeed the current induced torque originating from the spin orbit coupling of the system can trigger magnetization switching[9-18], coherent magnetization oscillation[19, 20], domain wall nucleation[21-24] and its motion[25-29].

At interfaces, spin orbit coupling is responsible for the perpendicular magnetic anisotropy[30, 31], which is essential for developing densely packed nanoscale magnetic elements. Technologically, the finding of perpendicular magnetic anisotropy at the CoFeB/MgO interface[30, 31] was particularly critical owing to the large tunnel magnetoresistance[32-35] the structure exhibits. In addition, recent reports have shown that the exchange coupling can also be modified by strong spin orbit coupling at interfaces. The Dzyaloshinskii-Moriya interaction[36, 37], an anti-symmetric exchange interaction which favors orthogonal alignment of neighboring magnetic moments, develops if the interface contains materials with strong spin orbit coupling[38]. In addition to the broken structural inversion symmetry of the system, strong interfacial DMI are responsible for the emergence of chiral magnetic structures, such as chiral domain walls[27-29, 39-46], spin spirals[47] and skyrmions[48-52].

Here we review and summarize our recent experimental results on the spin orbit effects in heavy metal (HM)/CoFeB/MgO heterostructures. Perpendicular magnetic anisotropy, the spin Hall effect, current induced torques and the Dzyaloshinskii-Moriya exchange interaction in heterostructures with different HM layers are studied. As these spin orbit effects are complementary and interrelated, we discuss individual effects as well as their correlations to provide a comprehensive view on this new frontier of Spintronics.

## 2. Experimental results

Films are deposited at room temperature using magnetron sputtering on Si (001) substrates coated with ~100 nm thick $SiO_2$. The film stack is: Si-sub|$d_{SEED}$ Ta|$d_N$ HM|$t_F$ $Co_{20}Fe_{60}B_{20}$|2 MgO|1 Ta (units in nanometers) where HM is one of the following transition metals: Hf, Ta, TaN, W and Re. To promote smooth growth of the HM layer, a Ta seed layer (thickness $d_{SEED}$) is occasionally formed before deposition of the HM layer. Reactive sputtering is used to form TaN: $N_2$ gas is added into the Ar gas atmosphere during the sputtering of Ta. The ratio ($Q$) of $N_2$ and Ar gas flows, measured using a mass flow meter attached to each gas line, is used to represent the $N_2$ concentration of the sputtering gas atmosphere. $Q$ is varied from 0, corresponding to pure Ta, to ~9% (see Ref. [53] for details). The resulting nitrogen composition of the film, evaluated using Rutherford backscattering spectroscopy (RBS), ranges between ~52 at% ($Q$~0.7%) to

~62 at% ($Q$~9%). All films are post-annealed at 300 ˚C for one hour in vacuum. No magnetic field is applied during the annealing process.

Magnetic moment ($M$) at saturation and the effective magnetic anisotropy energy ($K_{EFF}$) are measured at room temperature using vibrating sample magnetometry (VSM). $K_{EFF}$ is estimated from the areal difference between the out of plane and in-plane magnetization hysteresis loops. Positive $K_{EFF}$ corresponds to magnetic easy axis directed along the film normal.

Throughout this paper, we use the coordinate system defined as the following. Positive (negative) current flows along +x (−x), the film normal is parallel to the z axis and a right handed coordinate system is employed.

## 2.1. Perpendicular magnetic anisotropy

Figures 1(a)-1(e) show the HM layer thickness dependence of the saturated magnetic moment per unit volume ($M/V$) for various underlayers with $t_F$=1 nm (Refs. [29, 53]). Volume ($V$) of the film is calculated using the nominal thickness of the CoFeB layer ($t_F$) and the deposition area ($A$) of the film defined by the opening of the shadow mask used during the deposition process, i.e. $V=t_F·A$. Note that $M/V$ differs from the saturation magnetization ($M_S$) when a magnetic dead layer[53-55] forms within the magnetic (CoFeB) layer or when proximity induced magnetization (PIM) appears in the neighboring layer[56-58]. For the films studied here, we find evidence of non-zero magnetic dead layer thickness in many film structures but no indication of the PIM.

For HM=Hf and Ta, $M/V$ drops significantly as the HM layer thickness is increased beyond ~2-3 nm. The trend does not change when a Ta seed layer is inserted under Hf. $M/V$ remains nearly constant with $d_N$ when TaN is used for the HM layer[53]. In contrast to Hf and Ta, $M/V$ shows an increase with increasing HM layer thickness for HM=W and Re.

The HM layer thickness dependence of $K_{EFF}$ is shown in Figs. 1(f)-1(j) (Refs. [29, 53]). In most cases, $K_{EFF}$ varies with $d_N$ in a similar way $M/V$ does: $K_{EFF}$ drops upon increasing $d_N$ beyond ~2-3 nm for HM=Hf and Ta, it remains constant for HM=TaN and shows an uptrend as $d_N$ is increased for HM=W and Re. For the W underlayer films, $K_{EFF}$ abruptly decreases at $d_N$~5 nm for films without the Ta seed layer and at a much smaller thickness ($d_N$<4 nm) when the Ta seed layer is introduced. Such drop in $K_{EFF}$ is associated with a structural phase transition of W[59-62]: W forms an amorphous phase when its thickness is small but transforms to a highly textured body centered cubic (bcc) phase when $d_N$ exceeds a certain thickness.

A similar phase transition is also found for HM=Re (Ref. [62]). Re undergoes a transition from a thin amorphous-like phase to a thick highly textured hexagonal closed packed (hcp) phase at $d_N$~6 nm where changes in $M/V$ and $K_{EFF}$ are found. The highly textured hcp phase appears at $d_N$~6 nm for films with and without the Ta seed layer. From transmission electron microscopy studies[62], we also find that the surface roughness of the Re underlayer is much larger than the other elements: this is likely to do with the poor wetting of Re on $SiO_2$. It turns out that inserting the Ta seed layer can significantly improve the surface roughness[62], thus improving values of $M/V$ and $K_{EFF}$ at small $d_N$. For the thick Re underlayer films, the difference in $M/V$ and $K_{EFF}$ between the films with and without the Ta seed layer becomes negligible. With regard to the size of PMA, we find the Re underlayer films with the Ta seed layer shows larger $K_{EFF}$ than that of the thicker Re underlayer films, indicating that the thin amorphous-like phase serves as a better underlayer in promoting large $K_{EFF}$ against the highly textured hcp phase. This trend is similar to what has been observed for the W underlayer films.

The parameters that characterize the magnetic properties of the films are summarized in Fig. 2. We fit the $t_F$ dependence of $M/A$ with a linear function to obtain the saturation magnetization $M_S$ and the magnetic dead layer thickness $t_D$ from the slope and the x-axis intercept of the linear function, respectively[53]. The interface contribution to the magnetic anisotropy energy $K_I$ is estimated by fitting $K_{EFF}^*\cdot t_{EFF}$ vs. $t_{EFF}$ with a linear function and taking the y-axis intercept of the linear function, where $t_{EFF} \equiv t_F - t_D$ and $K_{EFF}^* \equiv K_{EFF} \cdot \frac{t_F}{t_{EFF}}$. $M_S$, $t_D$ and $K_I$ are plotted as a function of $d_N$ in Fig. 2.

For most cases $M_S$ (Figs. 2(a)-2(e)) lies in between that of the bulk $Co_{20}Fe_{60}B_{20}$ (~1500 emu/cm$^3$) and the corresponding CoFe phase without the boron, i.e. $Co_{25}Fe_{75}$ (~1890 emu/cm$^3$), both of which are represented by the horizontal dashed lines[63]. In general $M_S$ depends on the amount of boron present in the magnetic layer: if the boron can diffuse out from CoFeB, $M_S$ will increase and approach that of the CoFe phase. The amount of boron that diffuses out from CoFeB seem to depend on the material HM: $M_S$ becomes the largest when HM=Ta. This is consistent with earlier studies in which Ta is found to absorb boron from CoFeB[64-67]. There is also a possibility that the CoFeB layer intermixes with the neighboring HM layer, which will likely result in degradation of $M_S$. However, for most cases, $M_S$ does not fall below that of bulk CoFeB, indicating that the degree of intermixing is not large. This is also supported by elemental mapping studies using transmission electron microscopy (unpublished results).

The magnetic dead layer thickness (Figs. 2(f)-2(j)) shows a relatively large variation with $d_N$. We infer that the dead layer forms due to structural and/or chemical intermixing that occur at the HM/CoFeB interface (if this intermixing extends into the CoFeB layer, one would expect to see a reduced $M_S$). As the thickness of the dead layer is less than half a nanometer in most cases, it is difficult to identify the constituting elements. Recent studies[66] have shown that a thin TaB layer forms at the interface between Ta and CoFeB. It is yet to be confirmed whether the diffusion of boron is responsible for the formation of the dead layer for films with other HM underlayers.

Figures 2(k)-2(o) show the $d_N$ dependence of the interface magnetic anisotropy $K_I$. The $K_I$ values are in direct correspondence with the $K_{EFF}$ values shown in Fig. 1(f)-1(j). The largest $K_I$ of ~1.8 erg/cm$^2$ is found when HM=TaN[53]. To obtain a large $K_I$ in CoFeB/MgO based structures, it is considered important to remove the boron from the CoFeB/MgO interface and promote the Fe-oxygen bonding that is essential for establishing perpendicular magnetic anisotropy[58, 68-70]. The HM underlayer has to enable boron diffusion from the CoFeB layer; however too much diffusion may promote intermixing of HM and CoFeB, which will result in an increase of $t_D$ (and possible degradation of $M_S$). We infer that Ta$_{48}$N$_{52}$ ($Q$~0.7%) is at the right balance of promoting boron diffusion and simultaneously limiting the intermixing. Note that when more nitrogen is added to TaN, $t_D$ becomes nearly zero owing to the limited diffusion process (TaN is known to be a good diffusion barrier[71]). However, since the boron diffusion process also becomes limited, $K_I$ is reduced to nearly half of that shown in Fig. 2(m) (see

Ref. [53]). For HM=W and Re, Fig. 2(d) and 2(e) indicate that boron mostly remains within the CoFeB layer as $M_S$ is close to that of bulk CoFeB. As a consequence, $K_I$ is smaller than that of Ta and TaN.

## 2.2. Spin Hall effect

We use the spin Hall magnetoresistance[72-74] (SMR) effect to evaluate the spin Hall angle of the HM layer. The magnitude of SMR is proportional to the square of the HM layer spin Hall angle ($\theta_{SH}$) and thus provides a mean to evaluate $\theta_{SH}$ for various materials. Figures 3(a)-3(d) show the change in the resistance ($\Delta R_{XX}/R_{XX}^Z$) due to SMR as a function of $d_N$. $\Delta R_{XX}$ is the resistance difference when the magnetization of the magnetic (CoFeB) layer points along the $y$ ($R_{XX}^Y$) and the $z$ ($R_{XX}^Z$) axes, i.e. $\Delta R_{XX} = R_{XX}^Y - R_{XX}^Z$.

The inverse of $R_{XX}^Z$ is plotted as a function of $d_N$ in Figs. 3(e)-3(j) to estimate the resistivity of the HM layer. Data are fitted with a linear function in appropriate thickness ranges: the resistivity is proportional to the inverse of the linear function's slope. Fitting results are shown by the solid and dashed lines in Figs. 3(e)-3(j). Except for Ta, we find that the slope of $1/R_{XX}^Z$ vs. $d_N$ changes, in some cases quite abruptly, at certain thicknesses. These changes in the resistivity reflect the structural phase transition that takes places in the HM layer[59-62]. The corresponding structure of the HM layer, determined using cross-sectional transmission electron microscopy[62], is listed in each panel of Figs. 3(e)-3(h). For Hf, the growth mode changes at $d_N \sim 2$ nm: the thin region remains amorphous-like whereas the thicker region grows as hcp. In contrast, W and Re transforms its entire structure from amorphous-like to bcc and hcp, respectively, at $d_N \sim 5$- 6 nm. Ta (and TaN) remains amorphous for the thickness range studied. For all elements, we find that an amorphous(-like) phase is the dominant phase when the HM layer is thin. The resistivity of the amorphous-like phase for the four elements is shown in Fig. 4(a).

To estimate the spin Hall angle, we using a drift-diffusion model[75] to account for the SMR:

$$\frac{\Delta R_{XX}}{R_{XX}^0} \sim -\theta_{SH}^2 \frac{\lambda_N}{d} \frac{\tanh^2(d/2\lambda_N)}{1+\xi} \left[ \frac{g_R}{1 + g_R \coth(d/\lambda_N)} \right] \quad (1)$$

$$g_R \equiv 2\rho_N \lambda_N \operatorname{Re}[G_{MIX}]$$

where $\rho_N$ and $\lambda_N$ are the resistivity and the spin diffusion length of the HM layer, respectively. $G_{MIX}$ is the spin mixing conductance[76] that characterizes the absorption of the *transverse* spin current[77]. Here we assume the effect of the absorption of the *longitudinal* spin current[78], whose magnitude depends on the spin polarization (*P*) of the magnetic layer, is small. Transverse (longitudinal) spin current corresponds to flow of electrons whose spin orientation is directed orthogonal (parallel) to the magnetization of the magnetic (CoFeB) layer. $\xi \equiv (\rho_N t_F / \rho_F d_N)$ describes the current shunting effect into the magnetic layer: $\rho_F$ represents the resistivity of the magnetic layer.

The solid and dashed lines in Figs. 3(a)-3(d) (Ref. [62]) show fitted curves using Eq. (1). The parameters obtained from the fitting ($\theta_{SH}$ and $\lambda_N$) are summarized in Figs. 4(b) and 4(c) for the amorphous-like phase. If we were to take into account the effect of the longitudinal spin absorption[78], the spin Hall angle estimation will increase by ~10-20%. The spin Hall conductivity ($\sigma_{SH}$) is calculated using the following phenomenological relation:

$$\sigma_{SH} = \theta_{SH}/\rho_N \tag{2}$$

$\sigma_{SH}$ is plotted for the four elements in Fig. 4(d), which indicate that $\sigma_{SH}$ is strongly influenced by the number of 5$d$ electrons of the HM layer. The largest $\sigma_{SH}$ is found in W albeit its resistivity is the smallest among the four elements studied. The size of $\sigma_{SH}$ is similar to that of Pt reported previously[79]. The underlying mechanism of the spin Hall effect[80-86] in the amorphous 5$d$ transition metals, whether its origin is intrinsic or extrinsic, remains elusive. These results, however, indicate that the element dependent spin orbit coupling plays an important role in defining spin dependent transport and the spin Hall effect.

### 2.3. Current induced torque

Current induced torque that arises in thin film heterostructures due to application of in-plane current can be modeled using the framework developed for describing spin transfer torque[77, 87-89] in spin valve nanopillars and magnetic tunnel junctions. The Landau-Lifshitz-Gilbert (LLG) equation that includes both the damping-like and field-like spin transfer torques reads:

$$\frac{\partial \hat{m}}{\partial t} = -\gamma \hat{m} \times \left[ -\frac{\partial E}{\partial \vec{M}} + a_J \hat{m} \times \hat{p} + b_J \hat{p} \right] + \alpha \hat{m} \times \frac{\partial \hat{m}}{\partial t} \tag{3}$$

where $\alpha$ is the Gilbert damping constant, $\gamma$ is the gyromagnetic ratio, $-\frac{\partial E}{\partial \vec{M}}$ is the effective magnetic field that includes external, exchange, anisotropy and demagnetization fields and $\hat{p}$ is the spin direction of the electrons entering the magnetic layer ($\hat{p}$ points along the *y*-axis when the spin Hall effect generates spin current that diffuses into the magnetic layer). $a_J$ and $b_J$ represent the damping-like and field-like components of the spin transfer torque, respectively. $a_J$ and $b_J$ may depend on the relative orientation of $\hat{m}$ and $\hat{p}$ (see e.g. Refs. [90-92]). For simplicity, here we assume that they are constant.

We employ the adiabatic harmonic Hall voltage measurement technique[90, 93-95] to evaluate the current induced torque. An AC current (frequency ω) is applied to a Hall bar and the in-phase first harmonic Hall voltage (i.e. the fundamental mode at ω) and the out of phase second harmonic Hall voltage (the mode at 2ω) are measured using a lock-in amplifier. This technique probes the size and direction of the effective field that is associated with the current induced torque acting on the magnetic moments, i.e. it provides information of the damping-like effective field $a_J \hat{m} \times \hat{p}$ and the field-like component $b_J \hat{p}$. For films with magnetization easy axis directed along the film normal, such effective fields point along the film plane. An in-plane external magnetic field, directed along the current flow direction (along the x-axis) or transverse to it (along the y-axis), is applied during the harmonic Hall voltage measurements to evaluate the size of the effective field component along the external field direction. (Strictly speaking, the harmonic Hall voltage measured along one field direction by itself does not provide the corresponding effective field: one needs to measure the Hall voltages in both directions (along x- and y-axes) and use an analytical formula to calculate the two components of the effective field[95].) Note that the damping-like component depends on the magnetization direction whereas the field-like component is independent of it. Thus one can identify whether the effective field is damping-like or field-like by measuring its dependence on the magnetization direction.

The damping-like ($\Delta \vec{H}_{DL} \equiv a_J \hat{m} \times \hat{p}$) and the field-like ($\Delta \vec{H}_{FL} \equiv b_J \hat{p}$) components of the current induced effective field are plotted as a function of $d_N$ in Figs. 5(a)-5(d) and Figs. 5(e)-5(h), respectively (Ref. [29]). The effective field is normalized by the current density that flows in the HM layer. In many cases, both components of the effective field increase with increasing $d_N$ and tends to saturate. Such HM layer thickness dependence is expected for the torque that arise when the spin Hall effect is the source of the spin current[10, 96, 97]. Thus we infer that the predominant mechanism that generates the torque(s) is the spin transfer torque caused by the diffusive spin current from the HM layer generated via the spin Hall effect.

The thickness dependence of Hf and Re needs to be treated with care as the structure of the HM layer changes with $d_N$ (Ta and TaN remains amorphous for the thickness range studied here). For Hf, the growth mode changes from amorphous-like to hcp at $d_N \sim 2$ nm. Results from the spin Hall magnetoresistance measurements shown in Fig. 3(a) indicate that the spin Hall angle of the amorphous-like and hcp phases of Hf is similar. However, the effective field is smaller for the amorphous-like phase than that of the hcp phase: Fig. 5(a) and 5(e) shows that the effective field is nearly zero for the amorphous-like phase ($d_N \lesssim 2$ nm). This is partly due to the difference in the CoFeB magnetization (Fig. 1(a)) which influences the size of torque: the amorphous phase of Hf induces larger $M/V$ than that of the hcp phase. With regard to the thicker hcp-Hf underlayer films, it is not clear why the effective field tends to decrease as the thickness of the hcp phase increases. For the Re underlayer films, a structural phase transition takes place at $d_N \sim 6$ nm and we find a sharp change in the effective field there. Given the $d_N$ dependence of the SMR (Fig. 3(d)), the thin amorphous-like phase likely has a larger spin Hall angle than the thick hcp phase (the spin Hall angle of the hcp phase cannot be determined from the results shown in Fig. 3). In agreement with this view, the effective field of the amorphous-like phase is larger than that of the hcp phase.

The magnitude of the damping-like effective field is more or less in accordance with that of the spin Hall angle determined by the SMR measurements. Although the origin of the field-like component is under investigation, its HM layer thickness dependence is similar to that of the damping-like component regardless of the HM layer material. However, the relative magnitude of the field-like component to the damping-like component seems to depend on the HM layer. For example, the field-like component of the Re underlayer films is considerably smaller than that of films with HM=Ta and TaN albeit the similar size of the damping-like component. Note that the direction of the field-like effective field points opposite to the spin direction of the electrons entering the magnetic layer when the spin Hall effect takes place in the HM layer.. This seems to apply for many heterostructures[90, 91, 98] with a thin magnetic layer, many of which possess perpendicular magnetic anisotropy. Identifying the origin of the field-like torque as well as proper understanding of the spin mixing conductance remain one of the important issues that need to be addressed[99-102].

Finally we note that there remains an uncertainty in obtaining the current induced effective field using the harmonic Hall voltage measurements, which has been discussed in Ref. [29]. In general, when the planar Hall-like resistance becomes comparable to the anomalous Hall resistance due to the large contribution from the SMR, we find anomalies in the effective field. Such anomaly has been found in W underlayer films. To date, the origin of the anomaly is not clear. We have thus limited discussion of the effective field in systems which possess small planar Hall resistance. (The damping-like effective field can be estimated using the spin Hall angle, obtained for example by the SMR measurements, and the saturation magnetization; see Refs. [10, 103]).

**2.4. The interface Dzyaloshinskii-Moriya interaction**

The Dzyaloshinskii-Moriya (DM) interaction plays an essential role in current induced motion of domain walls driven by the spin Hall effect of the HM layer[104]. In order to utilize the damping-like torque exerted by the spin current generated via the spin Hall effect, the magnetization of the domain walls has to point along the current direction[104, 105]. The efficiency of the damping-like torque to drive the domain wall scales as $\sim a_J(\hat{m}_{DW} \times \hat{p})$, where $\hat{m}_{DW}$ is the direction of the domain wall magnetization. Since $\hat{p}$ is orthogonal to the current flow, $\hat{m}_{DW}$ must be pointing along the current flow to achieve non-zero efficiency. For wide wires (i.e. typically the width larger than ~100 nm) the magneto-statically preferred domain wall magnetization in perpendicularly magnetized films is the Bloch type[106], i.e. $\hat{m}_{DW}$ points transverse to the wire's long axis and is parallel to $\hat{p}$. However, the presence of HM layer with strong spin-orbit coupling can induce DM interaction at the interface to force the domain wall magnetization to point along the wire's long axis, i.e. to form Neel walls with a fixed chirality. When $\hat{m}_{DW}$ is orthogonal to $\hat{p}$, the efficiency of the damping-like torque takes its maximum and the domain wall moves at high velocity. Thus measuring the velocity of domain walls provides a mean to estimate the DM interaction[27-29, 107].

We have studied the magnitude and sign of the DM interaction at the HM/CoFeB interface using current induced motion of domain walls. In Fig. 6, we show representative results from HM=W. The CoFeB thickness dependence of the DM interaction is studied here. The magnetization per unit volume $M/V$ and the effective magnetic anisotropy energy $K_{EFF}$ are plotted as a function of the CoFeB layer thickness ($t_F$) in Figs. 6(a) and 6(b). The magnetic easy axis points along the film normal when 0.8 nm ≲ $t_F$ ≲ 1.2 nm.

The $t_F$ dependence of the propagation field ($H_P$) needed to induce motion of domain walls with the out of plane field is shown in Fig. 6(c). Although the variation of $H_P$ with $t_F$ is not large, the trend is similar to that of $K_{EFF}$, indicating that the domain wall width ($\Delta$), which scales with $\sim 1/\sqrt{K_{EFF}}$, plays a role in setting $H_P$. (See for example Ref. [108] for discussion of the relation between $H_P$ and the domain wall width.) The current driven domain wall velocity is measured as a function of the pulse amplitude and is plotted in Fig. 6(d) for different CoFeB thicknesses. The pulse amplitude is proportional to the current density that flows in the film. All devices exhibit domain wall motion along the current flow. Given the sign of the spin Hall angle of W[29, 61, 109], these results indicate that the domain walls in this system possess a right handed chirality. The thickness of the CoFeB layer has a non-negligible effect on the threshold pulse amplitude required to move the walls. Such variation can be explained by the changes in $M/V$ and $K_{EFF}$ with $t_F$, which influences the size of the damping-like torque and $\hat{m}_{DW}$.

The saturation velocity at large current can provide estimate of the DM exchange constant ($D$)[103, 104, 110]. Unfortunately, the films studied here are susceptible to current induced nucleation of domain walls which makes it difficult to observe the velocity saturation at large current. We have therefore used the relationship[27, 28] of velocity vs. in-plane external field directed along the current flow ($H_X$) measured at moderate current density to study $D$ as a function of $t_F$. As the DM interaction rotates the domain wall magnetization from the magneto-statically stable Bloch wall to that of a Neel wall, its effect can be modeled as an in-plane offset field ($H_{DM}$) pointing along the wire's long axis. Application of $H_X$ can thus cause the domain wall magnetization to rotate and change the velocity. The domain wall velocity as a function of $H_X$ is plotted in Fig. 6(e) for $t_F \sim 0.8$ nm. The velocity changes linearly with $H_X$ and is thus fitted by a linear function, as shown by the solid lines in Fig. 6(e). According to a one dimensional (1D) model of domain walls[27-29, 103, 104, 110, 111], the x-axis intercept of the linear line, defined as $H_X^*$, can be expressed as: $H_X^* = -\Gamma \left[ \frac{D}{M_s \Delta} + \text{sgn}(\theta_{SH}) \frac{2}{\pi} \frac{u}{\gamma \Delta} \right]$ (see Ref. [29] for the derivation), where $u$ represents the strength of the adiabatic spin torque term when current flows within the magnetic layer, and $\Gamma$ indicates the domain wall pattern ($\Gamma=-1$ for ↓↑ walls and $\Gamma=+1$ for the ↑↓ walls). Here for large $d_N$, majority of the current flows through the HM layer and thus contribution from the adiabatic spin torque term can be neglected, leading to $H_X^* \sim -\Gamma H_{DM} = -\Gamma \frac{D}{M_s \Delta}$. $H_X^*$ is positive (negative) for the ↓↑ (↑↓) walls, which corresponds to domain walls with right-handed chirality[112], i.e. $D>0$.

The $t_F$ dependence of $H_X^*$ is plotted in Fig. 6(f). The calculated DM exchange constant $D$ is plotted against $t_F$ in Fig. 6(g). Note that $D$ represents the DM exchange constant averaged across the entire thickness of the magnetic layer. Thus $D$ should scale with $1/t_F$ if the origin of DM interaction is at the interface[113]. To estimate the *interface* DM exchange constant ($D_I$), or the DM exchange constant per unit magnetic layer thickness, we plot $D$ vs. $1/t_F$ in Fig. 6(h) and fit the results with a linear function. As the structure of the W/CoFeB is predominantly amorphous, we assume a simple cubic lattice to calculate the atomic density within the layer. Assuming a lattice constant of ~0.28 nm, a value taken[63] from $Co_{25}Fe_{75}$, the slope of the linear function in Fig. 6(h) gives $D_I$~0.75 meV. This value is smaller than that reported[39] for W(001)/Fe, $D_I$=1.4 meV.

We find that the DM exchange constant strongly depends on the HM material used to interface the CoFeB layer. Figure 7 shows the average DM exchange constant $D$ of HM/CoFeB/MgO heterostructures with different HM layers[29, 114]. $D$ is negative for Hf, nearly zero for Ta, becomes positive for TaN and W. Note that the $D$ shown in Fig. 7 is for the thicker hcp phase of Hf, whereas the other three materials (Ta, TaN and W) are predominantly amorphous. We have also studied current induced motion of domain walls in Re/CoFeB with the Ta seed layer, however, we find domain walls hardly move with current, suggesting that $D$ is nearly zero since the damping-like torque is similar in magnitude with that of films with HM=Ta.

The results shown in Fig. 7 indicate that the DM exchange constant $D$ depends on the number of 5$d$ electrons, suggesting an electronic origin. As the change in the electronegativity with the number of 5$d$ electrons is similar to that of $D$ shown in Fig. 7, we infer that the relative difference between the electronegativity of the HM and CoFeB layer may play a role in defining the DM interaction. With regard to the origin of DMI[115, 116], it has been suggested that proximity induced magnetism (PIM) is responsible for the large $D$ at the Pt/Co interface[117]. Here we find little evidence of PIM in the HM/CoFe/MgO (HM=Hf, Ta, TaN, W, Re) heterostructures studied. As the size of $D$ of Pt/Co interface is at least a few times larger than that of HM/CoFeB, it remains to be seen whether PIM enhances $D$ of other systems.

It is well established that the DM exchange constant can be significantly modified by the structure (texture) of the interface[39, 118]. Whereas Pt/Co films with large $D$ typically possesses the fcc structure with (111) plane facing the interface[44, 119], the CoFeB in HM/CoFeB/MgO heterostructures is predominantly amorphous and many of the HM layers, including W which shows the largest $D$, is amorphous-like (the only exception is the thicker hcp Hf with which we find non-zero $D$), suggesting that contribution from the structure on $D$ may be small in this system. It should be noted that we find a non-zero thickness of a magnetic dead layer for the heterostructures with HM=W (see Fig. 2(i)). Since the constituents of the dead layer are not known at the moment, it is difficult to evaluate the influence of the dead layer on the DM interaction.

Finally, it has been reported recently that in Ta/CoFeB/MgO, the amount of boron present at the interface may influence $D$ via modification of local atomic configuration[120]. As the boron diffusion significantly depends on the material used for the HM layer (unpublished results), it is difficult to determine the degree of such contribution to the DMI for all the structures studied here. Further investigation is required to clarify quantitatively the origin of DMI at the HM/CoFeB interface.

## 3. Summary

We have described the effect of the heavy metal (HM) underlayer on the so-called spin orbit effects in CoFeB/MgO heterostructures. The perpendicular magnetic anisotropy at the CoFeB/MgO interface is significantly influenced by the HM underlayer. The boron diffusion process and intermixing at the HM/CoFeB interface are partly responsible for the changes in the saturation magnetization and the perpendicular magnetic anisotropy with the HM layer.

The spin dependent transport properties of the HM layer in the heterostructures are studied in connection to its structural phase. The spin Hall magnetoresistance (SMR), a powerful technique to characterize spin transport at the HM/magnetic layer interface, is utilized to estimate the spin Hall angle and the spin diffusion length. We find the spin Hall angle of the HM layer is considerably influenced by its structure. The amorphous-like phase possesses the largest spin Hall angle for the materials studied. Among the amorphous HM layers, we find that the spin Hall conductivity depends on the number of the $5d$ electrons each element possesses which is in accordance to what has been predicted for the intrinsic spin Hall effect.

The current induced torque is studied using the adiabatic harmonic Hall measurements. The changes in the damping-like component of the associated current induced effective field with the HM layer material and its thickness are consistent with the changes in the spin Hall angle and the spin diffusion length of the HM layer determined by the spin Hall magnetoresistance measurements. In almost all cases, the field-like component varies with the HM layer thickness in a similar way the damping-like component does, and its direction points opposite to the spin direction of electrons entering the magnetic layer if one assumes the spin Hall effect is the source of the spin current. The ratio of the field-like torque to the damping-like torque depends on the HM layer material. Further studies are required to identify the origin of the field-like torque as well as to provide proper understanding of the spin mixing conductance at the HM/magnetic layer interface.

Finally, the Dzyaloshinskii-Moriya (DM) interaction at the HM/CoFeB interface has been studied using current driven motion of domain walls in micron-size wires. The DM interaction varies with the HM layer material even though the CoFeB layer in many systems possesses a disordered amorphous-like structure. These results indicate that structural contribution to the DM interaction may be small in this system, and instead a strong electronic contribution may account for the origin. The largest DM interaction, ~0.75 meV per unit CoFeB layer thickness, is found for the W/CoFeB interface. Identifying the chemical and structural phase of the interface, in particular, its relation to the formation of the magnetic dead layer, is particularly important to describe the origin of the DM interaction in this system.

With the engineering of the film stack and materials innovation, it is very likely that new insights into spin orbit effects will continue to unveil for the coming years.

**Acknowledgement**

The authors thank J. Liu, T. Ohkubo, K. Hono, S. Mitani, S. Takahashi, S. Maekawa, M. Yamanouchi, S. Fukami, H. Ohno, E. Martinez, A. Thiaville, A. Kellock, S-H. Yang, S. Parkin for helpful discussions. This work was supported in part by Japan Society for the Promotion of Science (JSPS) KAKENHI Grant Numbers 25706017 and 15H05702, the Funding Program for World-Leading Innovative R&D on Science and Technology (FIRST Program) from JSPS and the MEXT R & D Next-Generation Information Technology.

**Figure captions**

**Figure 1.** Volume averaged saturation magnetization $M/V$ (a-e) and effective magnetic anisotropy energy $K_{EFF}$ (f-j) plotted against the heavy metal (HM) layer thickness. The film structures are Sub|$d_N$ HM|1 CoFeB|2 MgO|1 Ta (HM=Hf, Ta, Ta$_{45}$N$_{55}$, W, Re) (solid circles) and Sub|$d_{SEED}$ Ta|$d_N$ HM|1 CoFeB|2 MgO|1 Ta (HM=Hf, W, Re) (open circles). The HM layer used is denoted at the top of the corresponding panels. Part of the results are adapted from Ref. [29].

**Figure 2.** The heavy metal (HM) layer thickness dependence of the saturation magnetization $M_S$ (a-e), the magnetic dead layer thickness $t_D$ (f-j) and the interface contribution to the magnetic anisotropy energy $K_I$ (k-o). The film structures are Sub|$d_N$ HM|$t_F$ CoFeB|2 MgO|1 Ta (HM=Hf, Ta, Ta$_{45}$N$_{55}$, W, Re) (solid circles) and Sub|0.5 Ta|$d_N$ HM|$t_F$ CoFeB|2 MgO|1 Ta (HM=Re) (open circles). The HM layer used is denoted at the top of the corresponding panels. For HM=W (~3 nm thick), two sets of films were investigated and the data obtained for both sets are displayed in (d), (i) and (n).

**Figure 3.** (a-d) The heavy metal (HM) layer thickness dependence of the spin Hall magnetoresistance ($\Delta R_{XX}/R_{XX}^Z$). Solid and dashed lines show the fitting results using Eq. (1) for appropriate thickness ranges. (e-h) The inverse of the sheet resistance [$1/R_{XX}^Z/(w/L)$] versus the heavy metal layer thickness. (i,j) Expanded view of the plots shown in (g) and (h). Solid and dashed lines in (e-j) represent linear fitting to the data for appropriate thickness ranges. The thickness dependent structure of HM layer is noted at the top of panels in (e-h). The film structures are Sub|$d_N$ HM|1 CoFeB|2 MgO|1 Ta

(HM=Hf, Ta, W, Re) (solid circles) and Sub|0.5 Ta|$d_N$ HM|1 CoFeB|2 MgO|1 Ta (HM= Re) (open circles). Adapted from Ref. [62].

**Figure 4.** Resistivity $\rho_N$ (a), the spin diffusion length $\lambda_N$ (b), absolute values of the spin Hall angle $|\theta_{SH}|$ (c) and the spin Hall conductivity $\sigma_{SH}$ (d) of the heavy metals (HM) used in CoFeB|MgO heterostructures. The film structures are Sub|$d_N$ HM|1 CoFeB|2 MgO|1 Ta (HM=Hf, Ta, W) and Sub|0.5 Ta|$d_N$ HM|1 CoFeB|2 MgO|1 Ta (HM=Re). Results are shown for films with the HM layer possessing an amorphous-like structure. Results are taken from Ref. [62].

**Figure 5.** The damping-like $\Delta H_{DL}/J_N$ (a-d) and field-like $\Delta H_{FL}/J_N$ (e-h) components of the current induced effective field plotted as a function of the heavy metal (HM) layer thickness. The film structures are Sub|$d_N$ HM|1 CoFeB|2 MgO|1 Ta (HM=Hf, Ta, Ta$_{48}$N$_{52}$, Re) (solid symbols) and Sub|0.5 Ta|$d_N$ HM|1 CoFeB|2 MgO|1 Ta (HM=Re) (open symbols). The effective field is normalized by the current density ($J_N$) that flows into the HM layer. Squares and circles correspond to the effective field when the magnetization of the CoFeB layer is pointing along +z and –z, respectively. Part of the results are adapted from Ref. [29].

**Figure 6.** Volume averaged saturation magnetization $M/V$ (a), effective magnetic anisotropy energy $K_{EFF}$ (b) and the average out of plane field needed to move a domain wall, i.e. the propagation field $H_P$ (c), plotted against the CoFeB layer thickness ($t_F$). (d) The pulse amplitude dependence of the domain wall velocity. The CoFeB thickness of the films is denoted in the inset. (e) Domain wall velocity plotted as a function of in-plane field along the current direction ($H_X$) for $t_F$~0.8 nm. The domain wall type is ↓↑ (left panel) and ↑↓ (right panel). Blue squares and red circles indicate the wall velocity when positive and negative current pulses are applied, respectively. The pulse amplitude is ~10 V. (f) The in-plane field $H_X$ at which the velocity becomes zero ($H_X^*$) plotted as a function of $t_F$. (g,h) Dzyalohinskii-Moriya exchange constant $D$ as a function of $t_F$ (g) and $1/t_F$ (h). (f-h) Symbols indicate $H_X^*$ and $D$ for ↓↑ walls (circles) and ↑↓ walls (squares). $H_X^*$ and $D$ are determined by positive and negative currents; both results are plotted together. The solid line in (h) is a linear fit to the data. All results are from Sub|3 W|$t_F$ CoFeB|2 MgO|1 Ta. The wires used to evaluate domain wall motion are ~5 μm wide, ~40 μm long.

**Figure 7.** Dzyalohinskii-Moriya exchange constant $D$ estimated for Sub|$d_N$ HM|1 CoFeB|2 MgO|1 Ta. The center panel shows $D$ against the atomic concentration of N in TaN. The HM layer thickness dependence of $D$, which originates from the $d_N$ dependence of the magnetic parameters ($M/V$ and $K_{EFF}$), is fitted with a function derived from the 1D model (see Ref. [29] for the details). The error bars show the range of $D$ when contribution from spin transfer torque is changed: lower (higher) bound of the error bars corresponds to $P=0$ ($P=1$) and the symbols assume $P=0.7$[121]. Results are derived from Ref. [29] and taken from Ref. [114].

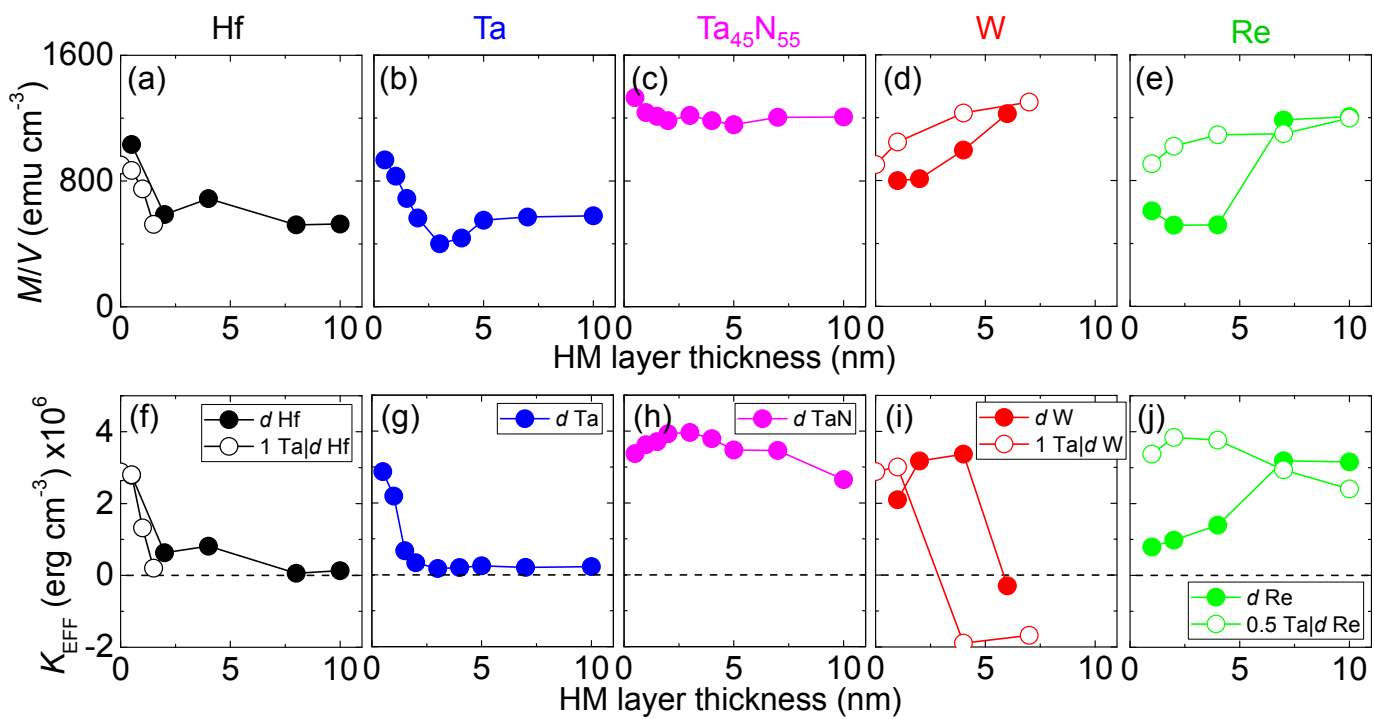

Fig. 1

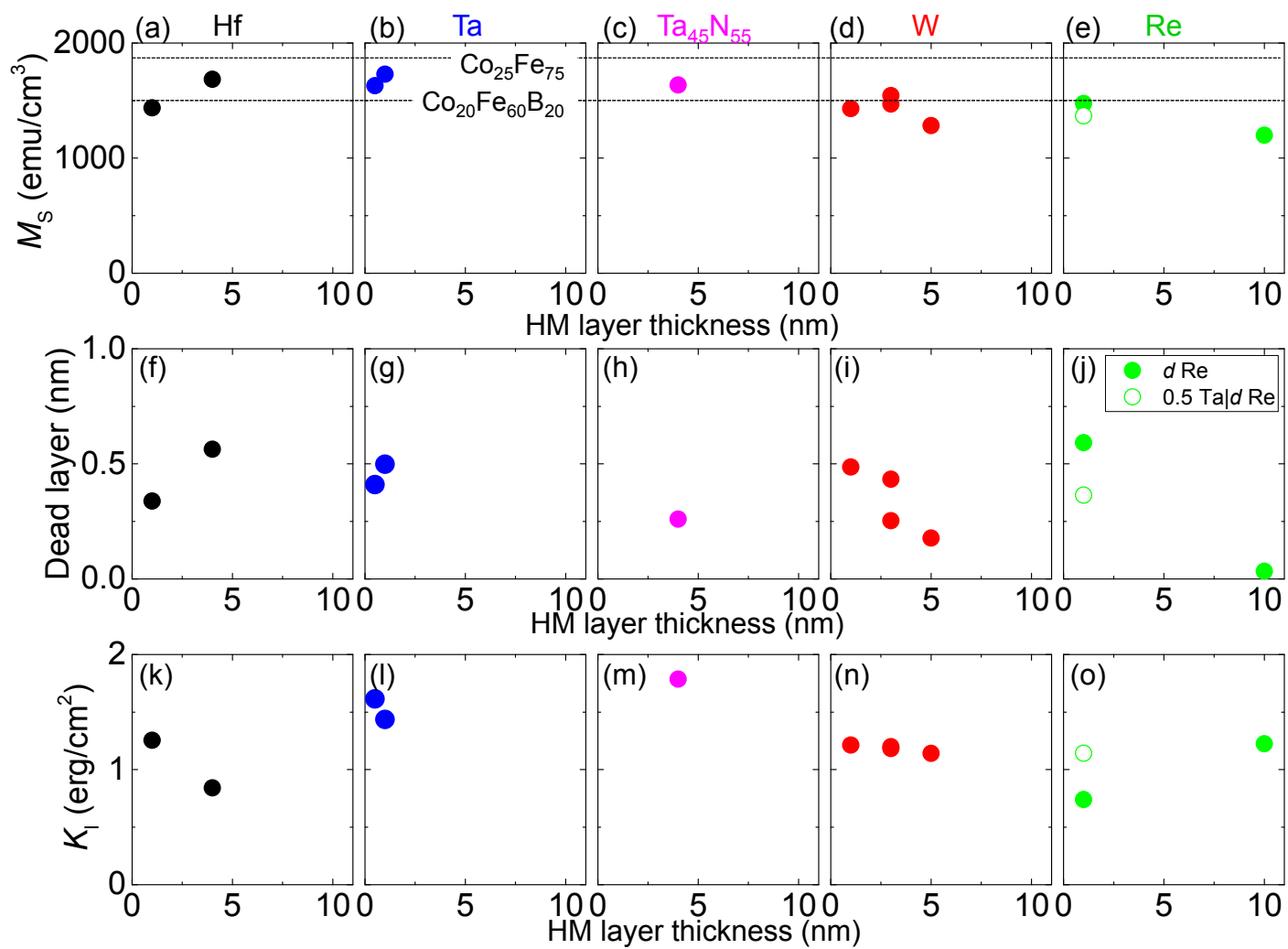

Fig. 2

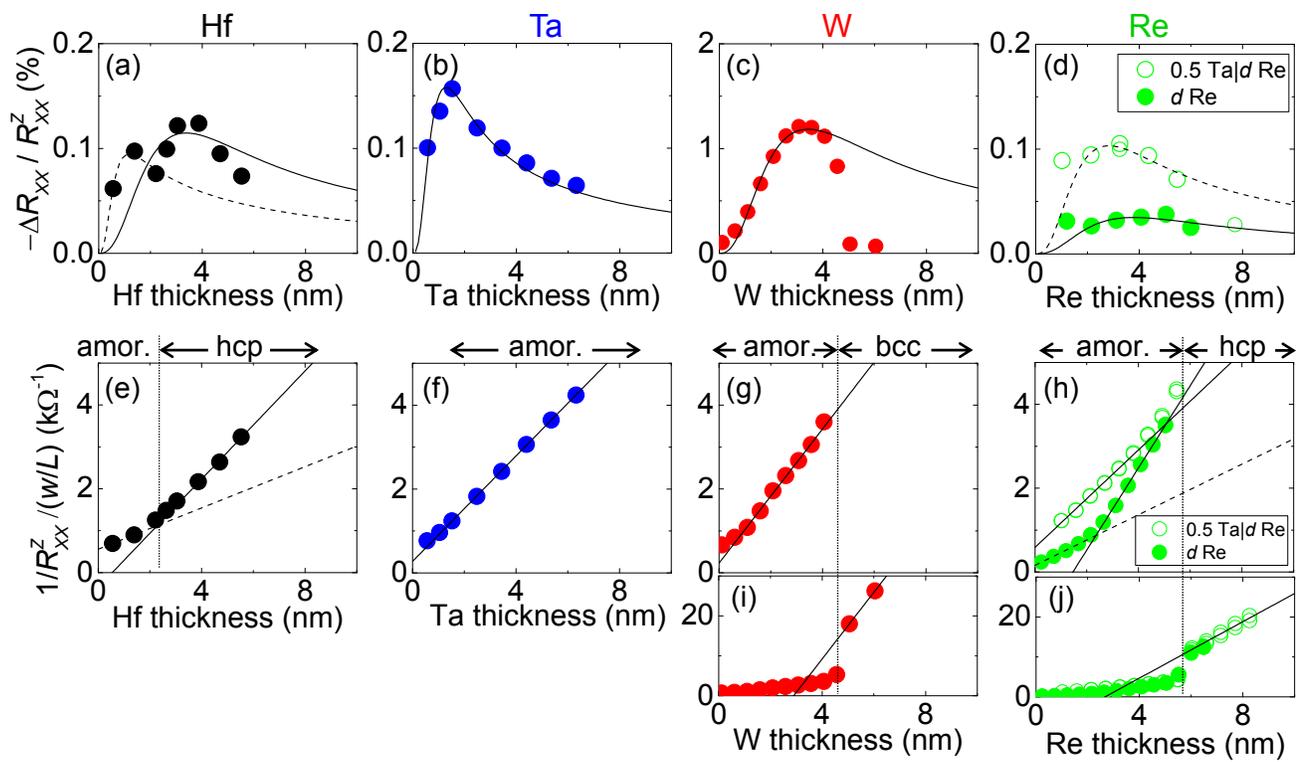

Fig. 3

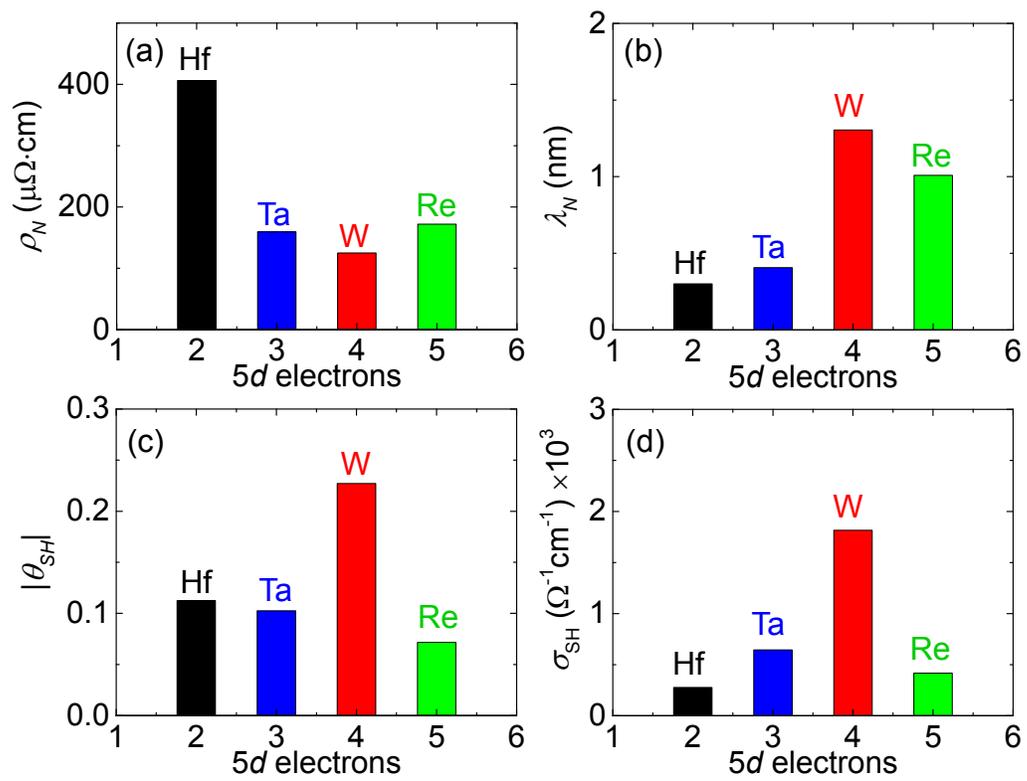

Fig. 4

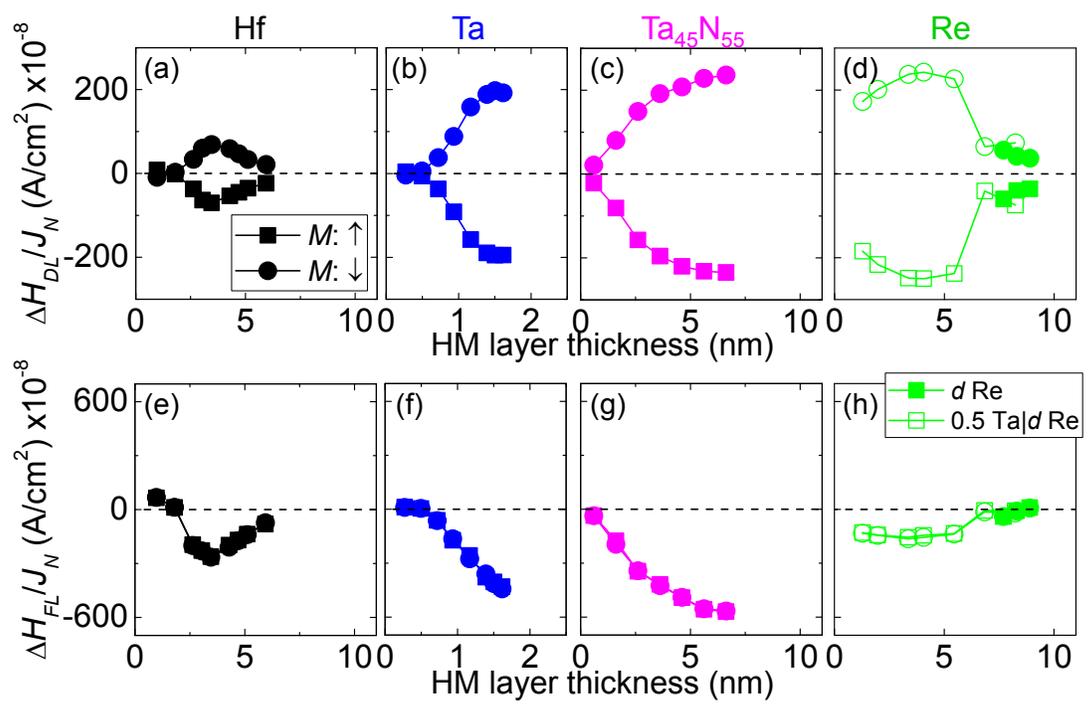

Fig. 5

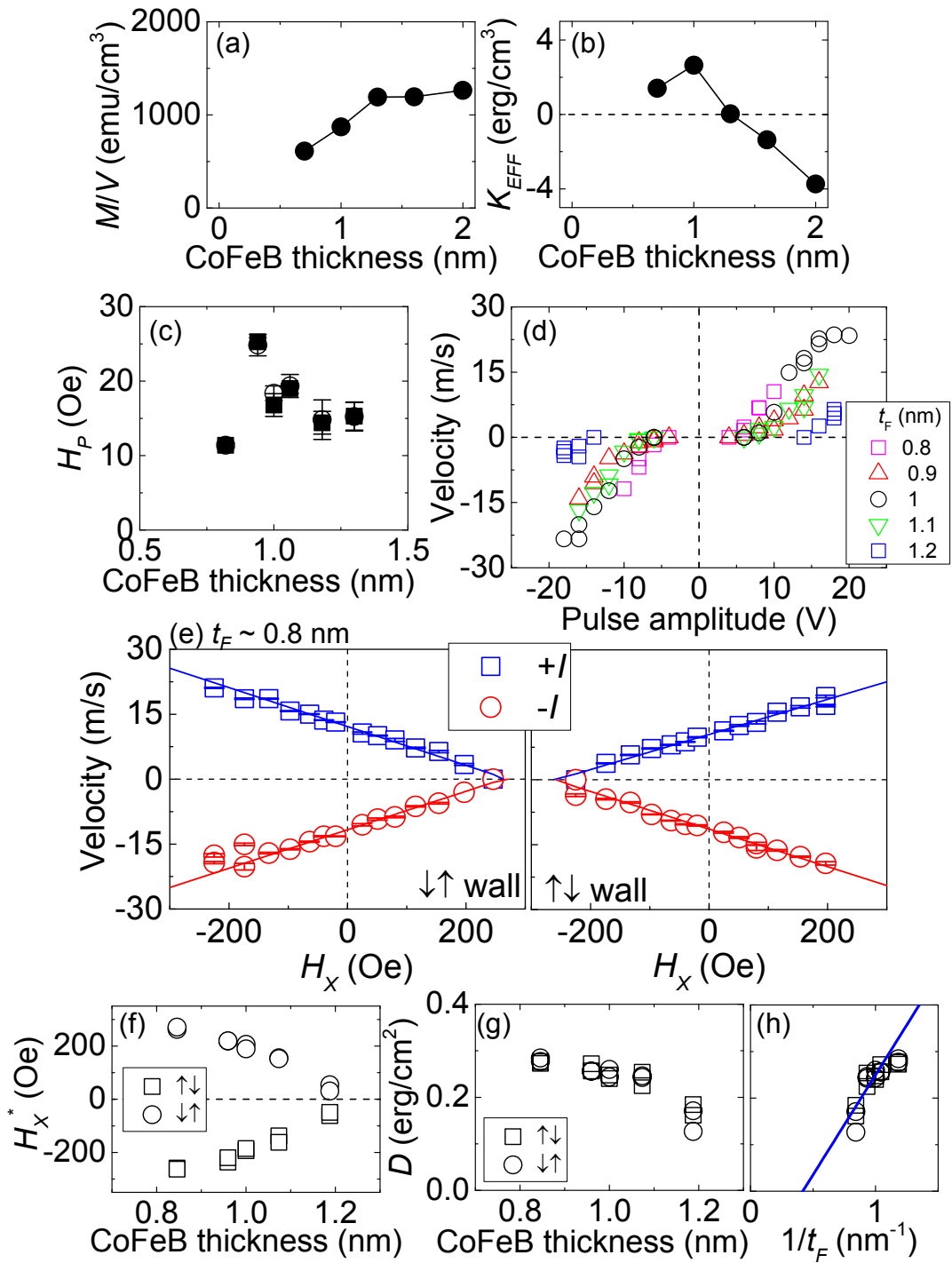

Fig. 6

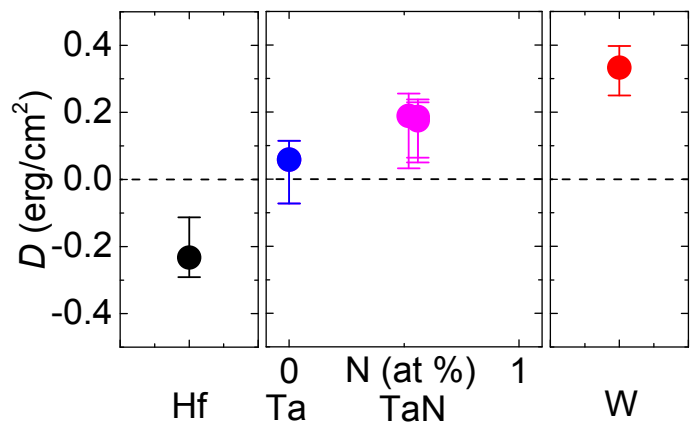

Fig. 7